\documentclass{aa} 

\usepackage{graphicx}
\usepackage{txfonts}
\usepackage{multicol}

\usepackage{xcolor}

\begin{document}

   \title{A quantitative  explanation of the radio - X-ray correlation in black-hole X-ray binaries}

   \author{Nikolaos D. Kylafis \inst{1,2}
          \and
          Pablo Reig \inst{2,1}
          \and 
          Alexandros Tsouros \inst{1,2}}
          
   \institute{University of Crete, Physics Department, University Campus, 70013 Heraklion, Greece\\
              \email{kylafis@physics.uoc.gr, pau@physics.uoc.gr, tsouros@physics.uoc.gr}
         \and
   Institute of Astrophysics, Foundation for Research and Technology-Hellas, 71110 Heraklion, Greece}

   \date{Received ...; accepted ...}

 
  \abstract
   {The observed correlation between the radio and X-ray fluxes in the hard state of black-hole X-ray binaries (BHXRBs) has been around for more than two decades now. It is currently accepted that the hard X-rays in BHXRBs come from Comptonization in the corona and the radio emission from the relativistic jet (Lorentz $\gamma \gg 1$), which is a narrow structure of a few $R_g=GM/c^2$ at its base.  The jet and the corona, however, are separate entities with hardly any communication between them, apart from the fact that both are fed from the accreting matter.  
   }
   {
   It is also widely accepted that the accretion flow around black holes in BHXRBs consists of an outer thin disk and an inner hot flow.  From this hot inner flow, which has a positive Bernoulli integral, an outflow must emanate in the hard and hard-intermediate states of the source.  By considering Compton up-scattering of soft disk photons in the outflow (i.e., in the outflowing "corona", which is a wider structure, tens to hundreds of $R_g$ at its base) as the mechanism that produces the hard X-ray spectrum, we have been able to explain quantitatively a number of observed correlations.  Here, we investigate whether this outflowing "corona" can explain the observed radio - X-ray correlation also.   We remark that the outflowing corona (wide, with low Lorentz $\gamma$) is completely separate from the relativistic jet (narrow, with high Lorentz $\gamma$). The two may co-exist, the jet at the rotation axis and the corona around it.
   }
   {
   We consider parabolic outflow models, that we have used successfully in the explanation of other correlations regarding GX 339-4 in the hard and hard-intermediate states, and compute the radio emission at 8.6 GHz coming from them, as well as the power-law photon-number spectral index $\Gamma$ of the Comptonized hard X-rays produced in them.  Thus, we have a correlation between the computed radio flux $F_R$ at 8.6 GHz and the computed spectral index $\Gamma$ of the hard X-ray spectrum.  This correlation is actually a {\it  theoretical  prediction}, since both $F_R$ and $\Gamma$ are computed from the model and, to our knowledge, no such correlation has been constructed from observations.  This prediction can be confirmed or proven wrong in future outbursts of GX 339-4.  From observations, we also produce a correlation between the observed hard X-ray flux $F_X$ and the observed index $\Gamma$.  Thus, for each value of $\Gamma$, observed or computed, we have the corresponding values of the observed $F_X$ and the computed $F_R$, which we plot one against the other.
   }
   {
   We find that, for GX 339-4, our model calculations for $F_R$ and $\Gamma$, with $\Gamma$ as the link between the observed $F_X$ and the computed $F_R$, reproduce extremely well the observed correlation of $F_R \propto F_X^{0.6}$ in the hard state.  In addition, in the hard-intermediate state of GX 339-4, this correlation breaks down and we {\it predict} that, in future outbursts of the source, the $F_R$ will exhibit first a sudden increase and then a sharp drop within a very narrow range of values of $F_X$.  Such a sharp drop of the $F_R$ has been observed in other sources.
   }
   {
   Since in our picture both the radio and the hard X-ray emission come from the same region, namely the outflow, it is not surprising that they are correlated. Also, since in a parabolic outflow with constant outflow speed the density is largest at its bottom, the soft photons, coming from below, see something like a "slab", with a moderate optical depth (up to 10 in the hard state) along the outflow and an order of magnitude larger in the perpendicular direction.  We remark that it is a slab geometry that is invoked to explain the observed X-ray polarization from BHXRBs.  Because of this, we {{\it predict}} that the X-ray polarization of GX 339-4 will be parallel to the outflow in the hard state and perpendicular to it in the hard-intermediate one.
   }

   \keywords{
                black-hole binaries --
                radio emission --
                X-ray emission --
                accretion --
                outflow --
                jet
            }

   \maketitle
%

\section{Introduction}

Black-hole X-ray binaries (BHXRBs) exhibit many observational correlations, but the most prominent one is the correlation seen in the hard state (HS) between the radio flux $F_R$, say at 8.6 GHz, and the X-ray flux $F_X$, say between 2 and 10 keV (Hannikainen et al. 1998; Corbel et al. 2000; 2003; Gallo et al. 2003, Bright et al 2020; Shaw et al. 2021).  In the so-called "standard track", the correlation is of the form $F_R \propto F_X^\beta$, where $\beta \approx 0.7$ (Gallo et al. 2003; Corbel et al. 2003) or $\beta \approx 0.6$ (Corbel et al. 2013).
Radio and X-ray emission occurs also in the hard-intermediate state, but the relation between the two has not been studied yet.

Extensive theoretical work has been done on both the radio and the X-ray emission from BHXRBs.  It is well accepted that the hard X-rays originate by inverse Compton scattering in the corona around the black hole (see, e.g., Done, Gierlinski, \& Kubota 2007; Kylafis \& Belloni 2015) and the radio in the relativistic jet (e.g., Fender, Belloni, \& Gallo 2004). Unlike wind-like outflows, the jet is considered to be a narrow structure, of the order of 10 gravitational radii at is base (e.g., Zdziarski, Tetarenko, \& Sikora 2022), with Lorentz factor $\gamma \gg 1$.
The jet and the corona are disjoint, but the jet is fed from the accretion flow/corona, and this led Heinz \& Sunyaev (2003) to examine scale invariant jet models, where they showed analytically that $F_R \propto (M \dot m)^{17/12}$, where $M$ is the mass of the black hole and $\dot m$ the accretion rate in units of the Eddington one.  For radiatively inefficient accretion flows, e.g. Advection Dominated Accretion Flows (ADAF, Narayan \& Yi 1994;1995; Abramowicz et al. 1995), for which $F_x \propto M {\dot m}^q$ with $q=2.3$, one has $F_R \propto M^{0.8} F_x^{0.6}$, which explains the observations very well (Merloni, Heinz, \& Di Mateo 2003).  If $q$ is somewhat different than 2.3, say $q=2$, the observed $F_R - F_X$ correlation cannot be explained.

A common picture of the accretion flow in BHXRBs in their hard and hard-intermediate states is that it consists of a geometrically thin, optically thick, outer disk (Shakura \& Sunyaev 1973) and a hot inner flow, ADAF, which is geometrically thick and optically thin (Narayan \& Yi 1994, 1995; Abramowicz et al. 1995).  This hot inner flow is taken as the corona in many models, including Heinz \& Sunyaev (2003) and Merloni, Heinz, \& Di Matteo (2003).  It is, unfortunately, not widely recognized that, due to the positive Bernoulli integral of the hot inner flow, a mildly relativistic outflow {\it must} emanate from it (Blandford \& Begelman 1999; Kazanas 2015).  In other words, {\it the Comptonizing corona is not static, but an outflowing one}.  This has tremendous implications.

 We have been promoting the idea that Comptonization in BHXRBs takes place mainly in the outflow (outflowing corona).  As we discuss in \S 4, our picture explains naturally a number of correlations, some of which have not been explained by any other model.  Due to the magnetic field that is needed for the ejection of the outflow (e.g., Blandford \& Payne 1982), the outflow emits also radio waves. Since it is the same electrons that do the Compton upscattering of the soft photons and the radio emission by synchrotron, it is likely that the radio and the X-rays are correlated.  This is what we demonstrate below. We restrict ourselves to GX 339-4, because it is well studied and we have modeled various correlations exhibited by this source (Reig \& Kylafis 2015; Kylafis \& Reig 2018, Kylafis, Reig, \& Papadakis 2020).
 
 We want to point out that we are not the only ones who have proposed an outflowing corona.  The first such proposal was made by Beloborodov (1999).   Also, other works (e.g., Malzac et al. 2001; Markoff 2005; Poutanen et al. 2023) have shown that the base of the outflow plays the role of the corona. We remark that the well-known JED-SAD model (Marcel et al. 2019; Barnier et al. 2022; see also previous work by this group), which is able to reproduce the spectral behavior (X-rays and radio) of BHXRBs during their outbursts, relies on self-consistent accretion-ejection solutions, unlike our simple model, which assumes the structure of the outflowing corona.

In \S 2 we describe our model, in \S 3 we present our results, and in \S 4 we discuss our findings.

\section{The model}

We perform Monte Carlo simulations of Comptonization in an extended outflowing region. We note that in our previous works, we referred to this outflowing region/corona as the jet, because in the past anything outflowing was called "jet". However, this name now appears
inappropriate. The outflow may or may not include a relativistic jet close to the black hole axis.
Our model consists of a parabolic outflowing corona with two symmetric lobes, where both Comptonization and radio emission occur.  We will describe the lobe along the positive $z$ axis.

\subsection{The outflowing corona}

Guided by the observations (Asada \& Nakamura 2012; Kovalev et al. 2020), we consider a parabolic outflowing corona whose radius at height $z>0$ above the black hole is given by
$$
R(z) = R_0 ~ (z/z_0)^{1/2},
\eqno(1)
$$
where $R_0$ is the radius of the outflow at its base, which is taken to be at a height $z_0$ above the black hole. Here, $R_0$ and $z_0$ are parameters of the model.

At the bottom of the outflow, there is an acceleration region, beyond which the flow speed is constant and equal to $v_0$.  Thus, the speed of the outflowing matter along the axis $z$ is taken to be
$$
v_{||} (z) = 
\begin{cases}
v_0 ~ (z/z_1)^a, & z_0 \leq z \leq z_1 \\
v_0, & z > z_1
\end{cases},
\eqno(2)
$$
where $v_0$, $z_1$, and $a$ are parameters of our model. 

If $n_e(z)$ is the number density of electrons along the outflow, mass conservation requires that 

$$
\dot{M} = 2 \pi R^2(z) m_p n_e(z) v_{||} (z) ,
\eqno(3)
$$
where $\dot{M}$ is the mass-outflow rate and $m_p$ is the proton mass. The outflowing matter is considered to be consisting of protons and electrons only. Eq. (3) implies that the number density of electrons along the outflow is given by
$$
n(z) = 
\begin{cases}
n_1 ~ (z_1/z)^{a+1}, & z_0 \leq z \leq z_1 \\
n_1 ~ (z_1/z), & z > z_1
\end{cases},
\eqno(4)
$$
where $n_1$ is the number density of electrons at $z_1$, while the density at the base of the outflow is $n_0= n_1 (z_1/z_0)^{a+1}$. The Thomson optical depth of the outflow along $z$ is given by 
$$
\tau_{||} = \int_{z_0}^{H}n_e(z) \sigma_T dz,
\eqno(5)
$$
where $\sigma_T$ is the Thomson cross section and $H$ is the height of the outflow.  Instead of $n_0$, we take $\tau_{||}$ as a model parameter.

In the rest frame of the flow, the electrons are generally taken to have a power-law distribution of Lorentz $\gamma$, namely
$$
N_e(\gamma_\text{co}) = N_0 ~ \gamma_\text{co}^{-p},
\eqno(6)
$$
from $\gamma_{min}$ to $\gamma_{max}$, where $p$, $\gamma_{min}$, and $\gamma_{max}$ are parameters of the model and for $N_0$ see below. Here, $\gamma_\text{co}$ is the Lorentz factor of the electrons in the \emph{co-moving frame}.

In order to calculate the distribution of the electrons for an observer at rest, one would need to perform the transformation of the Lorentz factor from $\gamma_\text{co}$ to $\gamma$. If one assumes that the velocity of the outflow is constant throughout, then it can be shown that $\gamma$ is approximately proportional to $\gamma_{co}$, and thus equation (6) also holds for an observer at rest. Since the acceleration region is small, the contribution of the acceleration region to the radio emission of the outflow is small.  Thus, we neglect the transformation from $\gamma_{co}$ to $\gamma$ and take the distribution of electrons for the observer at rest to be
$$
N_e(\gamma) \simeq N_0 \gamma^{-p}.
\eqno(6a)
$$

The normalization $N_0(z)$ can be calculated by integrating (6a) from $\gamma_{min}$ to $\gamma_{max}$, and equating the expression with the co-moving electron density. This yields

$$
N_0(z) = n(z) \sqrt{1-\frac{v_{||}^2(z)}{c^2}} (p-1) \gamma_{\text{min}}^{p-1}.
\eqno(7)
$$

\subsubsection{Parameter values}

As in our previous work, we fix all the parameters except for $R_0$ and $\tau_{||}$, which we vary.  The rest of the parameters have the values: $z_0= 5 R_g$, where $R_g=GM/c^2$ with $M$ the black-hole mass, $v_0=0.8c$, $z_1=50 R_g$, $a=1/2$, $p=3$, $\gamma_{min}=1.0$, and $\gamma_{max}=500$.  

Our results and conclusions are insensitive to reasonable values of the above parameters.  In particular, $v_0$ can be significantly less than 0.8 and all of our previous work (see \S 4) is unaffected (Reig \& Kylafis in preparation).  {\it The crucial ingredient of our model is the parabolic shape of the outflow}.  Non-parabolic outflows might be necessary to explain the, so called, outlier sources, but this will be the subject of a subsequent paper.

\subsection{Radio emission}

Ignoring synchrotron self-Compton (for a justification see Giannios 2005),
the equation for the transfer of radio photons in the outflow, 
in direction $\hat n$, along which length is measured by $s$, 
is given by
$$
{{dI(\nu, s)} \over {ds}} = j(\nu, s) - a(\nu, s)I(\nu,s),
\eqno(8)
$$
where $j(\nu,s)$ and $a(\nu, s)$ are the emission and absorption coefficients respectively, and $I(\nu, s)$ is the intensity at frequency $\nu$ at position $s$. The formal solution of this equation is 
$$    
I_\nu = \int_{s_1}^{s_2}ds j(\nu, s) \exp \left( -\int_s^{s_2} ds' a(\nu,s') \right). 
\eqno(9)
$$

For $\hat{n}$ perpendicular to the outflow axis, this is simplified to
$$
I_\nu (z)= \frac{j(\nu,z)}{a(\nu, z)} [1- \exp\{-a(\nu,z)R(z)\}],
\eqno(10)
$$
where $R(z)$ is given by eq. (1),
since the emission and absorption coefficients depend only on $z$. The total power radiated per unit frequency per unit solid angle is thus given by 
$$
\frac{d E_\nu}{dt d\nu d \Omega} = 2 \pi \int_{z_0}^H dz I_\nu(z) R(z). 
\eqno(11)
$$

Now, we need to specify the absorption and emission coefficients. Since the radio emission of the outflow is due to synchrotron emitting electrons, whose Lorentz factors follow a power law as in equation (6a),
the absorption and emission coefficients can be calculated analytically (Rybicki \& Lightman 1979). The expressions are

\setcounter{equation}{11}

\begin{align}
    a(\nu,z) &= \frac{\sqrt{3}q^3}{8 \pi m_e} \left( \frac{3q}{2 \pi m_e^3 c^5} \right)^{\frac{p}{2}}N_0(z)B^{\frac{p+2}{2}}(z)\\ \nonumber
    &\times\Gamma \left( \frac{p}{4}+\frac{1}{6}\right)\Gamma \left( \frac{3p+22}{12}\right) (m_e c^2)^{p-1}\nu^{-\frac{p+4}{2}},
\end{align}

and

\begin{align}
    j(\nu,z) &= \frac{\sqrt{3}q^3 N_0(z) B(z)}{4 \pi m_e c^2 (p+1)} \\ \nonumber
    &\times\Gamma \left( \frac{p}{4}+\frac{19}{12}\right)\Gamma \left( \frac{p}{4}-\frac{1}{12}\right) \left(\frac{ 2 \pi m_e c^2}{3qB(z)}\right)^{\frac{1-p}{2}} \nu^{\frac{1-p}{2}},
\end{align}
where $q$ and $m_e$ are the absolute value of the charge and the mass, respectively, of the electron, $B(z)= B_0 (z_0/z)$ is the strength of the magnetic field at height $z$ assuming flux freezing, $B_0=B(z_0)$, and $\Gamma$ is the Gamma-function.

Integrating equation (11) over solid angles, gives the power per unit frequency radiated, 

$$
P(\nu) \equiv \frac{d E_\nu}{dt d\nu} =  4 \pi^2 \int_{z_0}^H dz I_\nu(z) R(z). 
\eqno(14)
$$

Since there are two outflows with opposite directions to each other, an observer at a distance $d$, whose line of sight makes a 90-degree angle with the outflow axis, will measure a flux of

$$
F(\nu) = \frac{2P(\nu)}{4 \pi d^2},
\eqno(15)
$$
where $d$ is the distance to the source. Since the observer's line of sight is taken to be perpendicular to the outflow axis, there is no need to account for a Doppler shift.

\begin{figure}
  \includegraphics[width=\linewidth]{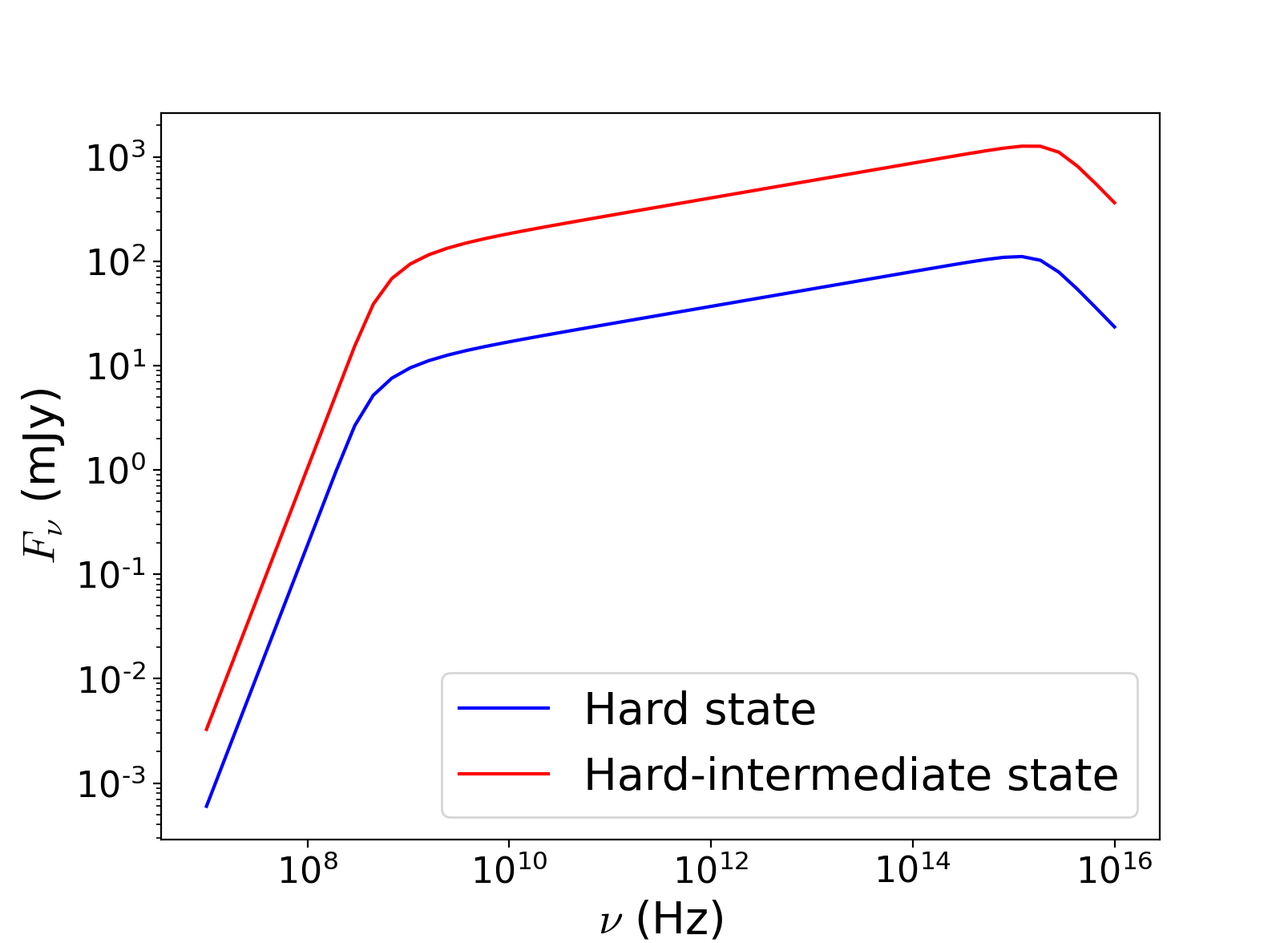}
  \caption{Flux per unit frequency produced
  by a power-law distribution of electrons with $p=3$ and $B_0 = 2 \times 10^5 $ G, for $R_0= 110 R_g$ and $\tau_{||}= 8.6$ (blue curve), and $R_0= 600 R_g$ and $\tau_{||}= 4.3$ (red curve). The power-law spectral index is $\alpha \approx 0.2$ in both curves.}
  \label{fig:1}
\end{figure}

In Fig. 1 we show the flux of the radio emission of the outflow for $p=3$, $B_0 = 2 \times 10^5$ G, and two values of the two main parameters, namely $\tau_{||} = 8.6$ and $R_0= 110 R_g$(blue line, corresponding to the hard state) and $\tau_{||} = 4.3$ and $R_0= 600 R_g$(red line, corresponding to the hard-intermediate state). The slope of the partially optically thick spectrum is $\alpha \approx 0.2$. 

\subsection{X-ray observations}
\label{xrays}

The X-ray data analyzed in this work correspond to the 2007-2008 outburst of GX 339-4 and were obtained with the Rossi X-ray Timing Explorer ({\it RXTE}). The data cover the interval MJD 53769--54678.
The details of the data analysis can be found in Reig et al. (2018) and Kylafis \& Reig (2018). 
We remark that the data and the models used in this paper are exactly the same as the ones used in Kylafis \& Reig (2018).
In this work, we restrict the analysis to the hard and hard-intermediate states during the rise of the outburst.   
We have obtained the energy spectra using the standard-2 mode of the
{\it RXTE}/PCA instrument in the energy range 3--9 keV to match the range used in Corbel et al (2013). We have fitted
the spectra with a broken power-law model and a Gaussian component that
represents the iron emission line at around 6.4 keV. We have allowed for
absorption at low energies. The hydrogen column density has been fixed
to $N_H = 4 \times 10^{21} \text{cm}^{-2}$  (Dunn et al. 2010). 

In Fig. 2, we show the observed correlation between the 3-9 keV X-ray flux and the photon-number spectral index $\Gamma$. Blue dots correspond to the hard state, while red dots to the hard-intermediate one.

\subsection{Comptonization}

In our model, Compton up-scattering of the soft input photons takes place in the outflow.  It is assumed that the soft photons come from the inner part of the accretion disk and have a black-body spectrum with $kT_{BB} = 0.2$ keV. This value of $kT_{BB}$ is not crucial.  The radiative transfer is done by our Monte Carlo code, which is in use for the past 20 years.  The soft photons are emitted isotropically upward at the base of the outflow. Typically, $10^7$ photons are enough to get good statistics.  The emergent hard X-ray spectrum is a power law with photon number index $\Gamma$.

In Kylafis \& Reig (2018), we used this Comptonization model and reproduced the observed values of $\Gamma$ in the hard and hard-intermediate states, along with the observed time-lags of the 9--15 keV with respect to the 2--6 keV photons.  {\it It is the same models that reproduce the correlation between time-lag and $\Gamma$ (Kylafis \& Reig 2018) that we have used to compute the radio spectra (see Fig. 1) and the radio flux at 8.6 GHz (used in Fig. 3).}

For each model (i.e., $\tau_{\parallel}$ and $R_0$) that produces with Comptonization in the outflow an index $\Gamma$, we compute the radio flux that comes out of the outflow.  
In Fig. 3, we show the pairs $(\Gamma,  F_R)$ from our model.  We remark that Fig. 3 is entirely theoretical.  In other words, it is a {\it quantitative, theoretical prediction} of our model for a future outburst of GX 339-4 (of course, also for previous ones, if simultaneous radio and X-ray data exist), similar to the one examined here. {\it Fig. 3 holds for the specific outburst studied here, yet we claim that the same curve (scaled up or down) should be observed in any outburst of GX 339-4}.  It is difficult to imagine how such a correlation can be produced by the typically considered model (hot inner flow plus  relativistic jet), since $\Gamma$ is fixed by the hot inner flow (corona) and the radio flux from the jet and the two do not seem to communicate with each other.  We remark further that Fig. 3 regards the rise of the outburst of GX 339-4.  We {{\it predict}} that in the declining part of the outburst, the curve will be traversed backwards, i.e., from high $\Gamma$ to low $\Gamma$, though the maximum radio flux will be lower than what it was in the rising part.

The radio-flux -- $\Gamma$ prediction above is related, and qualitatively similar, to the prediction made by Kylafis \& Reig (2018, see their Fig. 4) regarding the radio-break-frequency -- $\Gamma$ correlation in GX 339-4.  The present  prediction is probably easier to be confirmed or rejected than the earlier one.
As in Fig. 2, blue dots in Fig. 3 correspond to the hard state, while red dots to the hard-intermediate one.  

It is interesting to look at Figs. 2, 3, and 4 from a different point of view.  Fig. 2 is an observational correlation between $F_X$ and $\Gamma$.  Fig. 4 is an observational correlation between $F_X$ and $F_R$, where we are assuming, just for this discussion, that the sudden increase of $F_R$ and its subsequent sudden decrease in the hard-intermediate state (red points in Fig. 4) will be confirmed observationally. Therefore, it is an observational requirement that $F_R$ must be correlated with $\Gamma$.  In other words, Fig. 3 is an {{\it observational}} requirement that should be obeyed by all the models.  As we demonstrated above, Fig. 3 is obeyed by the outflow model.  It will be interesting to see if it is obeyed by the other models also.



\begin{figure}
   \centering
   \includegraphics[width=8cm]{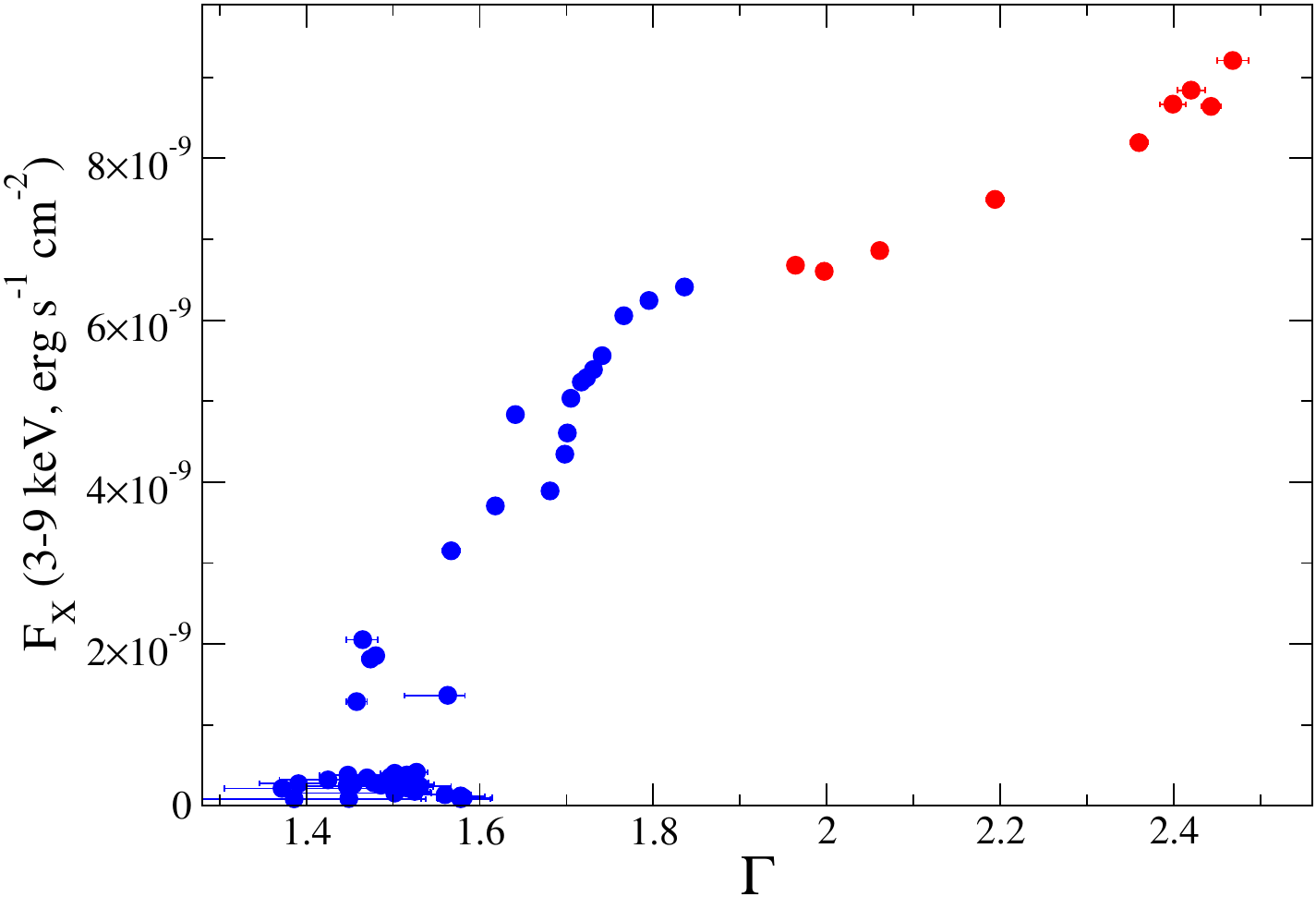}
      \caption{Relationship between the observed X-ray flux in the 3 -- 9 keV range and the photon index $\Gamma$.  Blue dots correspond to the hard state, while red ones to the hard-intermediate one.
              }
         \label{lx-gamma}
   \end{figure}


\begin{figure}
   \centering
   \includegraphics[width=8cm]{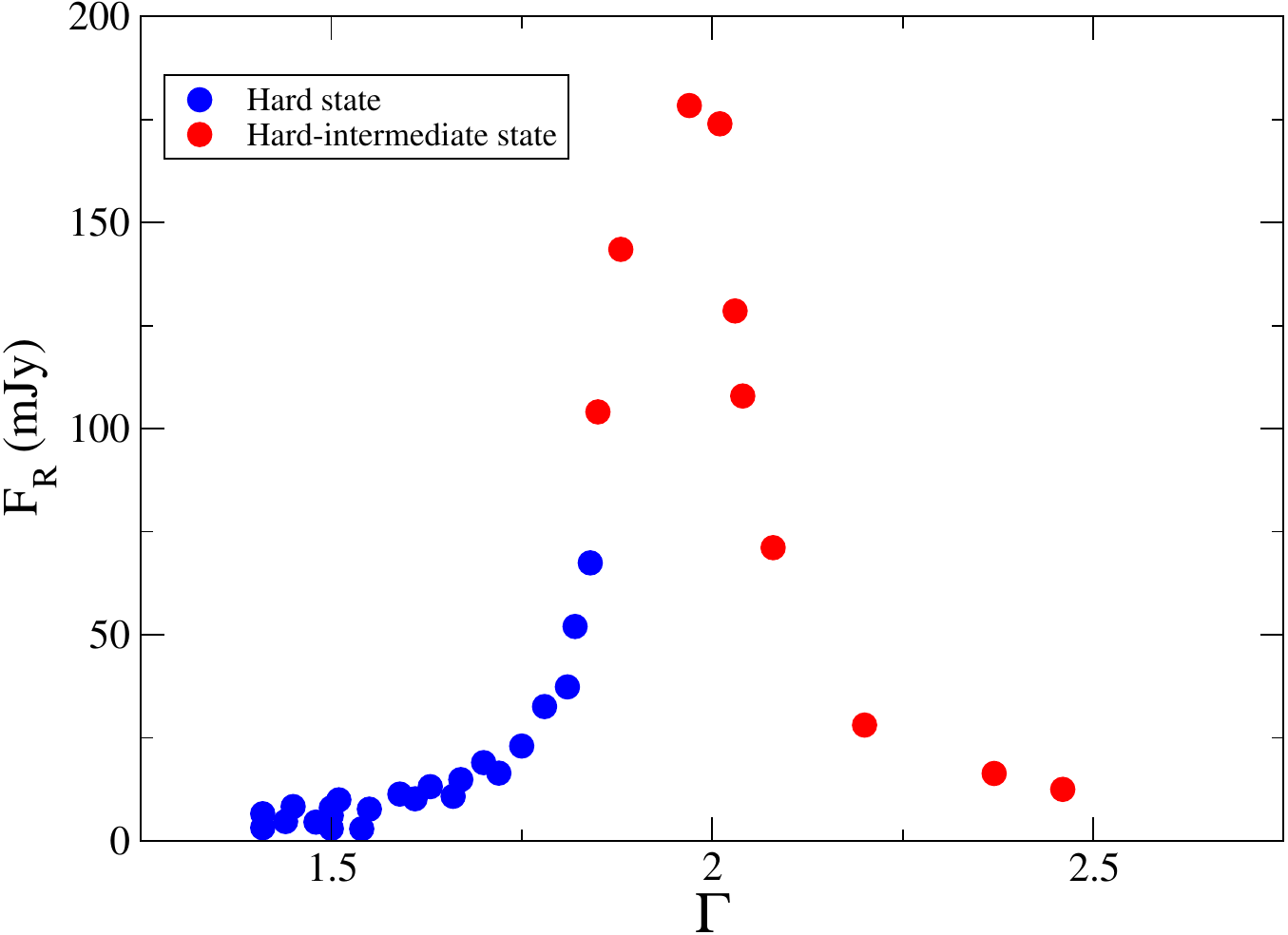}
      \caption{Relationship between the theoretical radio flux at 8.6 GHz and the photon index $\Gamma$.  This relationship is a theoretical prediction.  Blue dots correspond to the hard state, while red ones to the hard-intermediate one.
              }
         \label{lr-gamma}
   \end{figure}

\begin{figure}
   \centering
   \includegraphics[width=8cm]{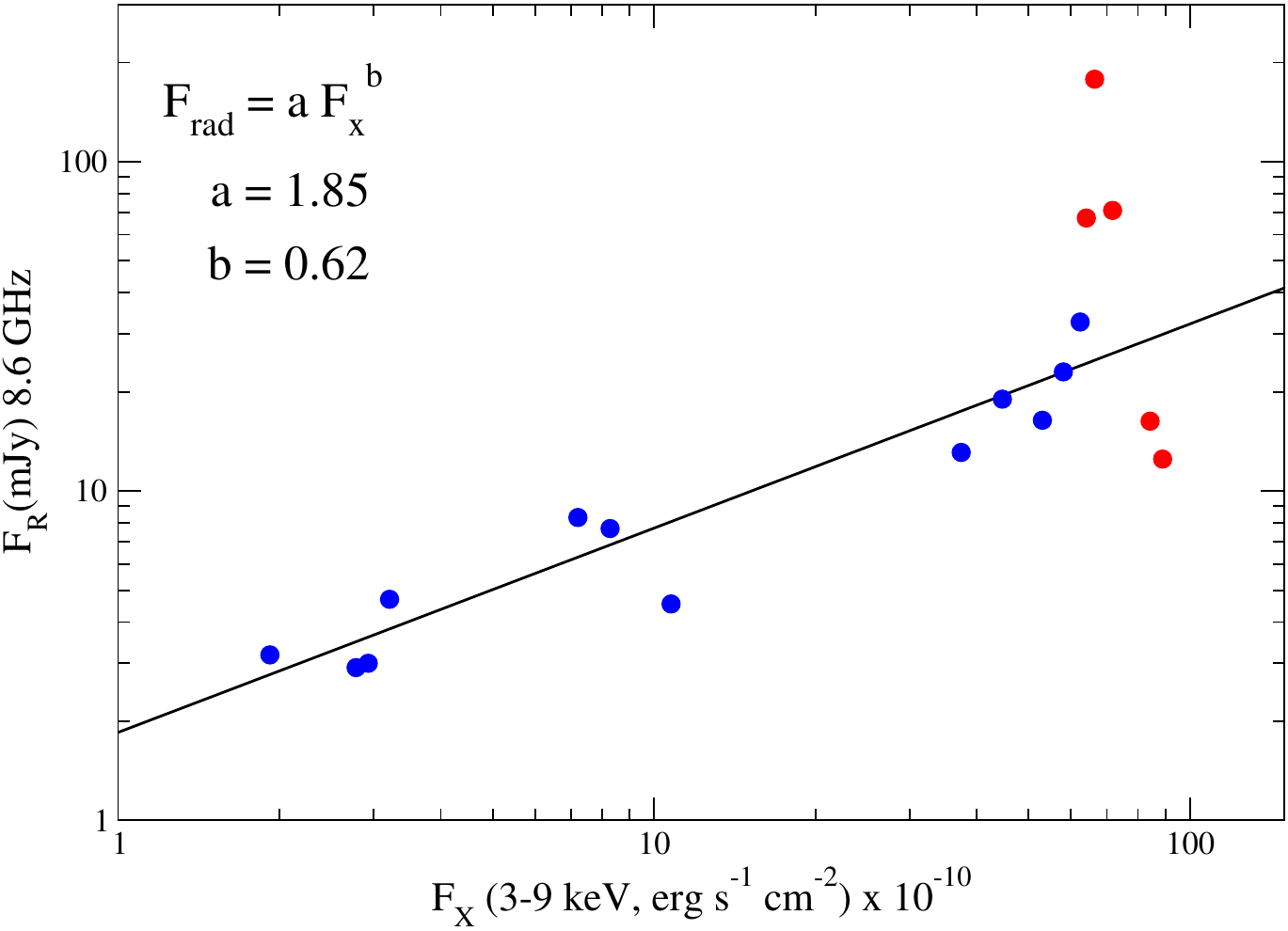}
      \caption{Relationship between the observed X-ray flux in the 3 -- 9 keV range and the computed radio flux at 8.6 GHz.  The black line is {\it not} a fit to the blue dots, but it is the observational correlation of Corbel et al. (2013).  Blue dots correspond to the hard state, while red ones to the hard-intermediate one.
              }
         \label{lx-lr}
   \end{figure}

\section{Results}

From Figs. 2 and 3, it is now clear that we can find the correlation between $F_R$ and $F_X$.  
We proceed as follows: we bin the X-ray data using the photon index $\Gamma$ in bins of size 0.02 for the hard state and of 0.1 for the hard-intermediate one. We compute the average value for each bin. Then we find the Monte Carlo model that produces a similar $\Gamma$ and we take the radio flux computed by it. In this way, we have pairs $(F_R, F_X)$ that correspond to the same $\Gamma$.


In Fig. 4, we plot the above found pairs $(F_R, F_X)$.  
As in Figs. 2 and 3, blue dots correspond to the hard state, while red dots to the hard-intermediate one. The black straight line {\it is not a fit to the blue dots}, but it represents the observational correlation found by Corbel et al. (2013). It is a power law function $F_R=a F_X^b$, where $a=1.85$ and $b=0.62$.  This power law breaks down when the source enters the hard-intermediate state.  There, the radio flux initially increases sharply and then decreases sharply within a narrow range of $F_X$ values (see also Fig. 3) and this is a {{\it prediction}} of our model.

The radio flux depends on the value of $B_0$ as $F_R \sim B^{(p+1)/2}$, where $p$ is the exponent in the electron energy distribution (eq. 6a).  Thus, we have selected the appropriate $B_0$ by shifting the data points only vertically, until they matched the observational correlation (solid line in Fig.~4).  In other words, the slope of the data points is the correct one.  The shifting of the points vertically is done by the parameter $B_0$.

The careful reader may have noticed that we have used eq. (10) for the calculation of the radio flux, because of its simplicity.  For an inclination angle between 40 and 60$^{\rm o}$, which is thought to be appropriate for GX339-4 (Shidatsu et al. 2011; Fürst et al. 2015), the calculation is much more complicated and the radio flux will be lower than what we have calculated.  The only effect that this will have is that the blue data points in Fig. 4 will not be shifted vertically as much and the red ones will not reach such high values of $F_R$.  In other words, the inferred value of $B_0$ will be slightly less than $2 \times 10^5$ G.

\section{Discussion}

We find it interesting that the observed slope of the $F_R - F_X$ correlation is reproduced so well with our outflow model.  As mentioned in the Introduction, a correlation between the two quantities is expected, because in the outflow model it is the same electrons that do the Comptonization and the synchrotron emission.  Of course, it could be any correlation and not necessarily the one observed.

In the hard-intermediate state, the fact that the radio flux drops sharply has been observed in a few sources (Gallo, Fender, \& Pooley 2003; Homan et al. 2005; Fender, Homan, \& Belloni 2009; Coriat et al. 2009; van der Horst et al. 2013; Russell et al. 2020).  Here, and for GX 339-4, we make the {\it prediction} that as the source moves from the hard to the hard-intermediate state, the radio flux will first {\it increase sharply} and then it will {\it decrease also sharply}.  Since the hard-intermediate state during the outburst rise lasts for only five days or so, high-cadence radio  observations are needed for the confirmation of this.

In this work, we have assumed that the observed radio emission comes from the outflowing corona.  It is not clear whether there is a narrow relativistic jet in the central part of the outflow.  If there is, our model makes the tacit assumption that it is not dominant.  We remark here that, in sources seen at high or moderate (like GX 339-4) inclination, the emission from a relativistic jet (Lorentz $\gamma \gg 1$) is expected to be weaker from the side than in the jet direction, because of Lorentz boosting. In such cases, the radio is expected to be dominated by the outflowing corona.

The question then naturally arises:  why sources whose radio emission comes mainly from the relativistic jet and sources whose radio emission is dominated by the outflow exhibit more or less the same $F_R - F_X$ correlation?  The answer seems to be that both the jet and the outflow are fed by the hot inner flow (ADAF).  Whether the Comptonization takes place in the corona, which is rotating around the black hole (hot inner flow, ADAF), or in an outflowing corona (outflow), the hard X-ray flux is more or less the same.  Similarly, whether the Blandford \& Payne (1982) mechanism is active and an outflow is produced, or the Blandford \& Znajek(1977) mechanism operates and a relativistic jet is produced, or both, the scaling relations of Heinz \& Sunyaev (2003) and Merloni et al. (2003) are valid.

Another question arises also: is the outflow model preferred to the typically considered model?  In our opinion, the answer is definitely yes, for the following reasons:  1)  Predictions. A model that, in addition to explaining the observations, also makes predictions, increases its credibility.  The outflow model makes two predictions:  that the radio flux should be correlated with $\Gamma$ (Fig. 3) and that the radio flux in the HIMS  must first increase suddenly and then decrease also suddenly (see red points in Fig. 3) as the source goes to the soft state. A related prediction was made in Kylafis \& Reig (2018) involving the radio break frequency and the photon-number spectral index.     
2) Simplicity. The outflow model is extremely simple (just Comptonization in and radio emission from a parabolic outflowing corona).  Nothing else! 3) Despite its simplicity, it explains quantitatively many correlations and observations, some of which have not been explained by any other model.  These are:

1) the energy spectrum from radio to hard X-rays for the source XTE J1118+480 (Giannios 2005),

2) the time-lag -- Fourier frequency correlation in Cyg X-1 (Reig, Kylafis, \& Giannios 2003),

3) the correlation between the time-lag and the photon-number spectral index $\Gamma$ in GX 339-4 (Kylafis \& Reig 2018) and other sources (Reig et al. 2018), 

4) the fact that this correlation depends on the inclination of the source (Reig \& Kylafis 2019),

5) the phase-lag – cutoff-energy correlation observed in GX 339-4 (Reig \& Kylafis 2015), 

6) the narrowing of the auto-correlation function with increasing photon energy seen in Cyg X-1 (Giannios, Kylafis, \& Psaltis 2004), 

7) the correlation between the Lorentzian frequencies in the power spectrum and the photon-number spectral index $\Gamma$ in Cyg X-1 and GX 339-4 (Kylafis et al. 2008),

8) the photon-number spectral index as a function of phase of the type-B QPO in GX 339-4 (Kylafis, Reig, \& Papadakis 2020),

9) the outflow provides a natural {\it lamppost} for the hard X-ray photons that return to the disk, where reflection and reverberation occurs (Reig \& Kylafis 2021), and

10) the $F_R - F_X$ correlation in GX 339-4 (this work).

11) Last but not least, the outflow model may explain naturally the observed X-ray polarization in BHXRBs (see the final remark below).

All of the above correlations are explained with only two parameters of the outflow: the radius $R_0$ at its base  and the Thomson optical depth $\tau_{||}$ along its axis.   For all the observations and correlations above, these two quantities vary in our model in the same narrow ranges $10 \buildrel < \over \sim R_0/R_g \buildrel < \over \sim 1000$ and $1 \buildrel < \over \sim \tau_{||} \buildrel < \over \sim 10$.  

{\it Final remark:} Due to the large density at the bottom of the parabolic outflow and its large size ($10$ to $10^3 R_g$), the soft input photons from the accretion disk, coming from below, see something like a "slab" at the bottom of the outflow, parallel to the accretion disk.  This is because for $\tau_{\parallel} \sim 10$ (appropriate for the hard state of GX 339-4) the height $h$, at which $\tau_{\parallel}$ from the bottom of the outflow is, say 3, is much less than the radius $R(z_0 + h)$ of the outflow there.  Therefore, $\tau_{\perp} \gg 3$ and the bottom of the outflow appears to the soft photons as a "slab". Compton scattering of X-rays in a slab parallel to the accretion disk produces polarization perpendicular to it, i.e., along the outflow (Beloborodov 1998).  This is what has been observed with IXPE (Krawczynski et al. 2022).  If our picture is correct, then we {{\it predict}} that the X-ray polarization direction in GX 339-4 will change  from parallel to the flow to perpendicular to it, as the source moves from the hard state ($\tau_{\parallel} \sim 10$) to the hard-intermediate one ($\tau_{\parallel} \sim 1$).

\begin{acknowledgements}
      We would like to thank Pierre-Olivier Petrucci for very useful comments on an earlier version of this paper and Iossif Papadakis, Andrei Beloborodov, Tom Russell, and Greg Marcel for useful discussions.  Also, N.D.K acknowledges useful e-mail exchanges with Rob Fender and David Russell regarding radio observations in the hard-intermediate state.
\end{acknowledgements}

%
%

\end{document}